\documentclass[runningheads]{llncs}
\usepackage[T1]{fontenc}
\usepackage{graphicx}

\usepackage{fancyhdr}
\usepackage[normalem]{ulem}
\usepackage[hyphens]{url}
\usepackage[sort,nocompress]{cite}
\usepackage[final]{microtype}
\usepackage[keeplastbox]{flushend}
\usepackage[bookmarks=true,breaklinks=true,letterpaper=true,colorlinks,citecolor=blue,linkcolor=blue,urlcolor=blue]{hyperref}

\usepackage{cite}
\usepackage{amsmath,amssymb,amsfonts}
\usepackage{algorithmic}
\usepackage{graphicx}
\usepackage{textcomp}
\usepackage{xcolor}
\usepackage{balance}

\usepackage{tabularx}
\usepackage{makecell}

\usepackage{hyperref}
\hypersetup{hidelinks}

\usepackage{pgfplots} 
\usetikzlibrary{patterns}

\title{FPGA-extended General Purpose\\Computer Architecture
}%

\author{Philippos Papaphilippou\inst{1}\orcidID{0000-0002-7452-7150} \and
Myrtle Shah\inst{2}\orcidID{0000-0002-5410-1646}
}

\institute{Department of Computing, Imperial College London, UK\footnote{The first author is now with Huawei Technologies R\&D (UK) Limited.}\\
\email{p.papaphilippou17@alumni.imperial.ac.uk}\\
\and
ChipFlow Ltd, UK%
}

\authorrunning{P. Papaphilippou et al.}

\begin{document}
\maketitle

\pagestyle{plain}

\begin{abstract}

This paper introduces a %
computer architecture, where part of the instruction set architecture (ISA) is implemented on small highly-integrated field-programmable gate arrays (FPGAs). %
Small FPGAs inside a general-purpose processor (CPU) can be used effectively to implement custom or standardised instructions. %
Our proposed architecture 
directly address related challenges for high-end CPUs, where such highly-integrated FPGAs would have the highest impact, such as on main memory bandwidth. %
This also enables software-transparent context-switching. %
The %
simulation-based evaluation of a dynamically reconfigurable core shows promising results %
approaching the performance of an equivalent core with all enabled instructions. %
Finally, the feasibility of adopting the proposed architecture in today's CPUs is studied through the prototyping of fast-reconfigurable FPGAs and studying the miss behaviour of opcodes. 
\end{abstract}

{\small \keywords{computer architecture \and memory hierarchy \and reconfigurable extensions}}

\section{Introduction}

{\center
\vspace{-16.5cm}
\hspace{-1.5cm}\mbox{Accepted at the 18th International Symposium on Applied Reconfigurable Computing (ARC) 2022}
\vspace{15.5cm}
}

There has been considerable maturity %
around traditional software on today’s CPUs. This has led to easier development through high-quality libraries and debug tools, as well as relatively mature programming models and verification routines. Additionally, a variety of software and hardware abstractions have enabled portability of code, such as with virtual memory and cache hierarchies, and enabled more effortless increase in performance, such as through instruction-level parallelism. %

However, general purpose processors leave a lot to be desired in terms of performance, hence the increase in use of computation offloading to specialised processors. These include graphics processing units (GPUs), FPGAs and even purpose-built silicon in the form of application-specific integrated circuits (ASICs). %

One consideration in today’s hardware specialisation technologies is the fact that they are mostly based on the non-uniform memory-access model (NUMA). Large off-chip memories are found in the majority of today’s high-end FPGA offerings, resulting in power-hungry and expensive setups, as well as in limitations in programming models, and complicating deployment and data movement.

While promising techniques like wide single-instruction multiple-data (SIMD) instructions \cite{intel} in CPUs %
attempt to close the gap between specialised and general purpose computing \cite{bordawekar2010can}, this gap is wider than it has ever been. This is because of the increased need for highly customised architectures in trending workloads \cite{wang2019benchmarking}, whose functionality cannot be efficiently expressed with a fixed general purpose ISA and architecture.

In this paper, we extend the most common computer architecture in today's systems (modified Harvard architecture \cite{grosbach2004modified}) to introduce FPGA-based instruction implementations in general purpose systems. In contrast to current research, this goes beyond embedded and heterogeneous processors, and introduces multi-processing for operating systems and fine-grain reconfiguration, as with standardised instruction extensions. %
A feasibility study shows promising performance for supporting reconfigurable extensions on-demand, especially when supporting fast FPGA reconfiguration. %
The list of contributions is as follows:
\begin{enumerate}
	\item The ``FPGA-extended modified Harvard Architecture'', a novel computer architecture to introduce FPGAs working as custom instructions, enabling context-switching and other advanced concepts for higher-end applications.
	\item A comprehensive evaluation with fine-grain reconfiguration (at the instruction-level), providing %
	insights %
	on the impact of the reconfiguration time and the operating system's scheduler properties for multi-processing.
	\item Feasibility studies elaborating on the readiness of current SoC technology to adopt the proposed  approach %
\end{enumerate}

\section{Challenges}\label{chall}

The research on FPGAs implementing instructions can be considered an attempt to overcome a series of challenges in current systems. %
This work addresses challenges found in existing %
research on custom instructions. 

\paragraph{\textbf{Current CPUs and Discrete FPGAs.}}\label{curcf}

One challenging design choice that relates to both hardware and software %
is %
the selection of instructions that would be more beneficial to include as part of the instruction set architecture (ISA). With a fixed ISA, vendors can select a subset of instructions, such as with the modularity of RISC-V \cite{waterman2020risc}, or design custom instructions. %
For general purpose computing it is difficult to predict what the most appropriate instructions will be. For instance, some applications may be ephemeral, as with some deep learning models, for which specialised hardware becomes obsolete faster.

Another challenge is hardware complexity. Supporting a high-number of instructions is expensive, but sometimes this has been unavoidable for widening the applicability of general purpose processors. For example, Intel's AVX2 and AVX-512 include thousands of instructions \cite{intel}, and RISC-V's unratified vector extension hundreds \cite{rvv}. The related implementation complexity, such as with AVX-512, is associated with a decrease in operating frequency and power efficiency, and area increase \cite{gott,cebrian2020scalability}. Additionally, AVX-512 is suboptimal for certain workloads, where a serial code could surpass them in terms of performance and scalability \cite{cebrian2020scalability}. %
Expanding the ISA can also harm the SoC scalability to many-cores, which heavily relies on core miniaturisation %
and power efficiency.

When using FPGAs as accelerators, one of the most limiting bottlenecks to performance is the bandwidth to main memory \cite{psurvey}. %
For example, even with Intel's Xeon+FPGA, %
although the FPGA is directly connected to the memory controller it only achieves 20 GB/s \cite{alonso2019doppiodb}. The memory hierarchy tends to always favour CPU performance, hence the presence of expensive off-chip memories in high-end FPGA boards. This heterogeneity is considered to impact FPGA development and increases the cost and deployment of FPGAs in the datacenter \cite{simodense}.

\paragraph{\textbf{FPGAs Implementing 
Instructions.}}\label{curef}

The basic limitation of the related work on FPGA-based instructions is the focus on embedded and/or heterogeneous systems, with no notion for multi-processing, context-switching and other advanced micro-architectural features. The use of embedded FPGAs (eFPGAs) has many practical applications in embedded systems \cite{ahmed2011efpgas,koch2021fabulous}, but there is currently no computer architecture to ``hide'' reconfiguration from traditional software.

One challenge in existing methods of introducing FPGAs as custom instructions is the need for manual intervention for reconfiguration. Although the recommended procedures to %
handle bitstreams can be well documented, deviating from conventional software development could be detrimental for %
 adoption \cite{ahmed2011efpgas}.

By initially focusing on highly-customised instructions and more complex accelerators, %
there has been less opportunity for modern processors to gradually adopt small reconfigurable regions as part of their core. It is more complex to derive conclusions from specific custom instructions and accelerators, as their exploration usually shifts the focus to specialisation and optimisation. %

\section{Solution}\label{sols}

\begin{figure}[h!]
\centering
\includegraphics[width=1.0\textwidth, trim=0 0 0 0]{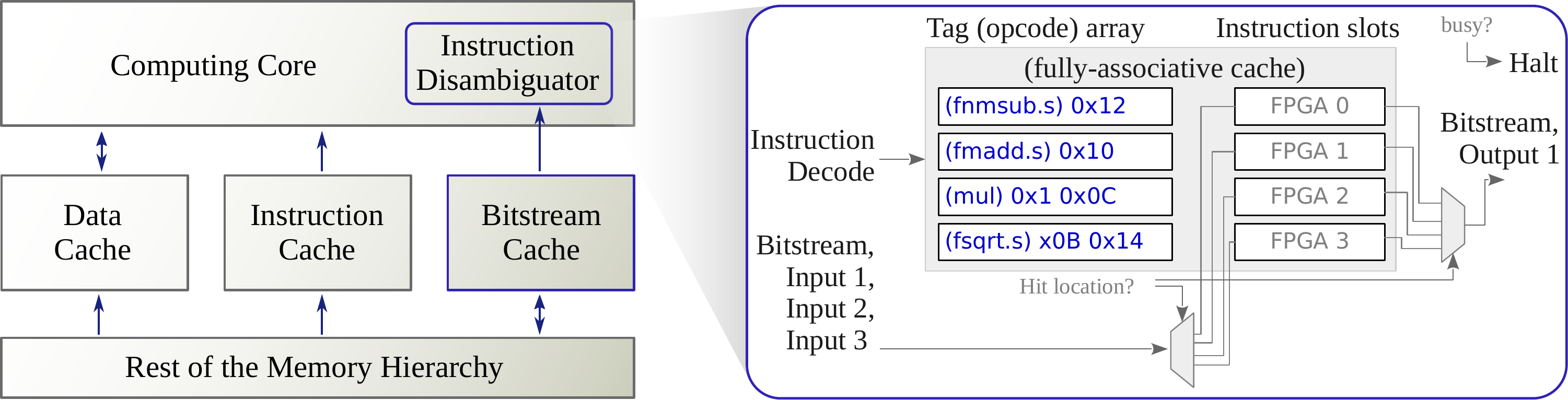}
\caption{%
Proposed computer architecture (left) and instruction disambiguator (right) }\label{harv}
\end{figure}

The proposed solution is the ``FPGA-extended modified Harvard Architecture'', which unifies the address space for instructions, data, as well as for FPGA bitstreams. When compared to the traditional modified Harvard architecture, the proposed solution also adds a separate bitstream cache at level 1, to provide bitstreams for FPGA-instructions after an instruction opcode is ready. %
The idea is for a computing core %
that features reconfigurable slots for instructions, to be able to efficiently fetch instruction bitstreams transparently from the software. %
Figure \ref{harv} (left) introduces the proposed computer architecture. %

This architecture assumes that the computing core features fast-programmable %
FPGAs that can be used to implement instructions. This can be achieved with the help of a small cache-like structure, the \emph{instruction disambiguator}, shown %
in figure \ref{harv} (right). On every instruction decode there is a request to this unit to see if there is an instruction implementation for the requested instruction. It operates as a fully-associative cache and uses opcodes (plus any additional fields for defining functions) as tags to determine the bitstream locations. On an opcode miss, it requests the instruction bitstream from the bitstream cache, while on a hit it multiplexes the operands to the appropriate slot.

The \emph{bitstream cache} is a separate cache specifically designed for FPGA bitstreams that can increase the performance of the reconfigurable core. %
Similarly to today's modified Harvard architecture, the L1 instruction and data caches are still separated and connected to a unified cache, allowing easier simultaneous memory accesses for pipelining the instructions. Since the instruction disambiguator unit waits for an instruction opcode to be ready, a bitstream fetch phase can be placed subsequently to the instruction decode pipeline stage in heavily-pipelined processors. This cache is separated to also allow different features than the rest of the caches, such as with wider blocks to facilitate the increased width to carry bitstreams, as opposed to instructions (see section \ref{bsds}). %

This approach enables the applications to be agnostic of the reconfiguration aspect. An operating system can provide ISA extensions (or part of them) in the form of bitstream libraries, while %
the hardware fetches the corresponding %
bitstreams on demand. Sharing the same %
address space for the bitstreams also enables keeping bitstreams in software binaries, so that they can provide custom instruction extensions alongside their data segment for %
acceleration potential.

\section{Evaluation}\label{evals}
This evaluation works as a proof of concept, thus any platform limitations are not %
handed-down over the proposed computer architecture.
As our proposal concerns a fundamentally different computer architecture and targets high-end hardened processors, future research on a detailed evaluation would involve fabrication.
The framework for evaluating the performance of the proposed architecture\footnote{%
Source code available: \url{https://github.com/pphilippos/fpga-ext-arch}
} is based on Simodense \cite{simodense}, an open-source FPGA-optimised RISC-V softcore, which was heavily modified to facilitate our study on the system effects of our proposal. %
This study extends it with the ``F'' extension for single-precision floating-point support. This resulted in the RV32IMF, where ``I'' is the base 32-bit integer and ``M'' is the integer multiplication/division extension \cite{waterman2020risc}.
Most of the ``I'' instructions introduce one cycle of latency, while the ``M'' instructions occupy 4 non-blocking cycles of latency. The ``F'' extension is pipelined with a latency of 6 cycles, excluding the fused multiply-add instructions that yield a 12-cycle latency. RISC-V's ``Zicsr'' and a set of control status registers (\emph{mstatus, mie, mcause, mepc} and \emph{mtvec}) were also added to support the experiment of section \ref{miben}.

The main addition is the instruction %
disambiguator. %
Its functionality here is to process opcodes (and related fields) and add artificial latency when there is an instruction slot miss (or hit). All required instructions actually pre-exist on the softcore, emulating the performance overhead of the proposal as observed by the software. The instruction opcodes are first being resolved through the instruction cache, and the instruction slot disambiguator here works as an L0 instruction cache that uses opcodes as cache tags and adds latency on opcode misses. %

With respect to the size and complexity of the reconfigurable instructions, we explore a compartmentalisation scenario, where instructions are grouped into single reconfigurable regions according to their logic similarity.  %
There are 3 groups for the ``M'' extension 
 (\{\emph{mul, mulh, mulhsu, mulhu}\}, \{\emph{div, divu}\}, \{\emph{rem, remu}\}), and 7 groups for the ``F'' extension
\{\emph{fadd.s, fsub.s}\},
\{\emph{fmul.s}\},
\{\emph{fdiv.s}\},
\{\emph{fsgnj.s, fsgnjn.s, fsgnjx.s, fmin.s, fmax.s, fle.s, flt.s, feq.s}\},
\{\emph{fsqrt.s}\},
\{\emph{fcvt.w.s, fcvt.wu.s, fcvt.s.w, fcvt.s.wu}\},
\{\emph{fmadd.s, fmsub.s, fnmsub.s, fnmadd.s}\}), totalling 10 groups. The number of free slots is parameterisable.

This emulates an environment where the CPU has no space for %
all extensions, and the workload exhibits competitiveness for a limited number of instruction slots. This instruction selection and granularity is indicative, thus a more complete %
ISA research would be appropriate to decide what fraction of instructions remains hardened in final products. Such an exploration would relate to the features and performance of the embedded FPGAs, while still allowing custom extensions.

The resulting codebase is synthesisable and also passed benchmark-based test cases on a Xilinx Zynq UltraScale+ FPGA. However,  as the resulting framework ran relatively fast using Verilator 4.224%
, we opted to use simulations instead. %

\subsection{Benchmark Classification}\label{fispe} %

The utilised benchmark suite is Embench \cite{patterson2020embench}, providing a %
selection of benchmarks with different attributes of interest. %
It was ported for use in our infrastructure, and each benchmark was made to run as a thread instead of a process. This required some additional modification, such adding thread safety for shared local libraries. %
Some benchmarks with double-precision floating point arithmetic were modified to use single-precision to make use of the ``F'' extension. %

In order to quantify the impact of the %
studied extensions on the benchmark performance, %
they are first seen individually. 
There are four binaries/runs per benchmark, one for each of the following fixed specification combinations: RV32I, RV32IF, RV32IM and RV32IMF.  %
When a useful instruction is absent from the specification of the compiler, it is replaced by a sub-optimal pre-defined routine, as specified by the %
application binary interface (ABI). %
The underlying softcore supports their superset RV32IMF and can run all 4 binaries per benchmark.

\begin{figure}[h!]
\centering
\includegraphics[width=0.9\textwidth, trim=0 5 0 0]{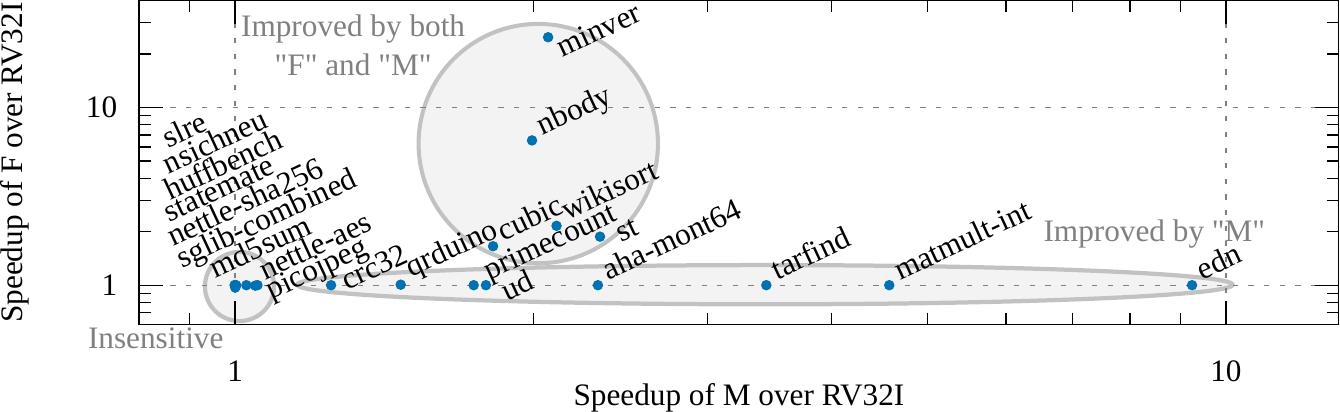}
\caption{Task classification based on the speedups of RV32IM and RV32IF over RV32I}\label{ficlass}
\vspace{-0.4em}
\end{figure}

The benchmark classification is %
illustrated in figure \ref{ficlass}. The axes represent the speedup of using one of ``M'' or ``F'' over only the base instruction set RV32I. As expected, the five benchmarks that used floating point all seem benefit from ``F'' (\emph{minver, wikisort, st, nbody} and \emph{cubic}), while ``M'' seems a relatively more popular set amongst the benchmark selection (\emph{crc32, qrduino, primecount, ud, aha-mont64, tarfind, matmult-int} and \emph{edn}). The remaining 9 benchmarks are classified as ``insensitive'', which exhibit different properties such as being control-heavy. Interestingly, there is no class where an Embench benchmark is only benefited from ``F'' and not from ``M'' here. %

\subsection{Single-program}\label{siben}%

For the evaluation of the proposed architecture under single benchmarks, we select the ``improved by both F and M'' class from the classification of the previous section. This is done to focus on workloads where there is demand for both instruction extensions, before introducing multi-processing. 

In this experiment with simulated reconfigurability, there are six data series for different miss and hit latency combinations for the instruction slot disambiguator. 
There are 10-cycle, 50-cycle and 250-cycle miss latencies representing both reconfiguration technologies that approach a latency closer to that of CPU instructions, and slower which could be achievable with more traditional partial reconfiguration techniques. %
For each of the three, there are versions with and without a hit latency, which is useful to represent potential discrepancies between the CPU core and the fabric (e.g. frequency drops).

\begin{figure}[h!]
\centering
\includegraphics[width=1\textwidth, trim=0 5 0 10]{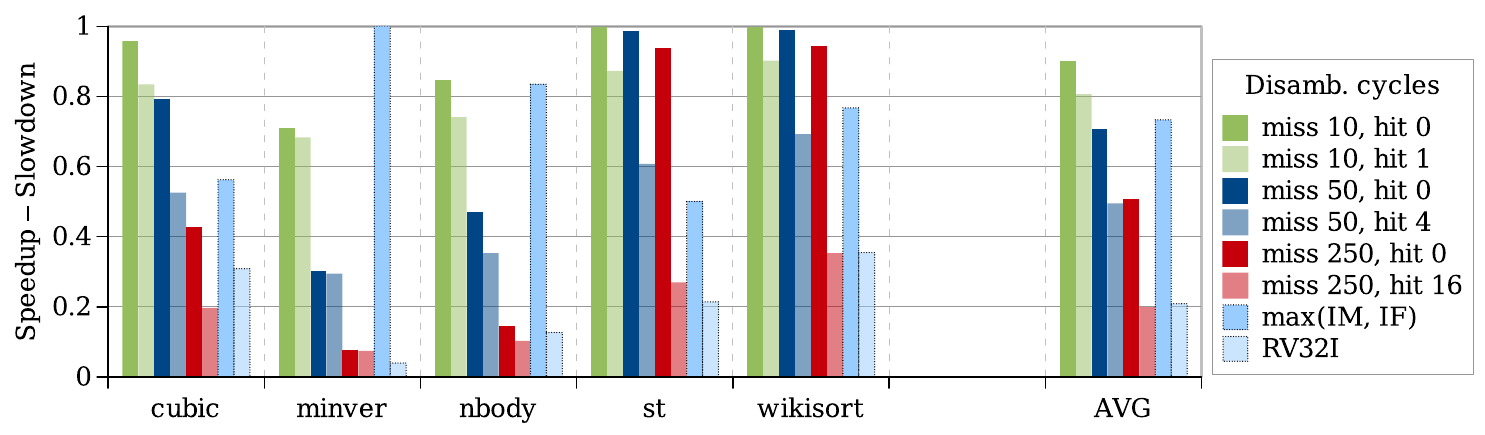}%
\caption{Approaching RV32IMF with reconfigurability for single benchmarks}\label{fireco}
\vspace{-0.4em}
\end{figure}

Figure \ref{fireco} presents these results for 4 available instruction (group) slots. %
The y axis shows the slowdowns %
over when running with a fixed specification with both ``M'' and ``F'' (RV32IMF). Note that all series regard slowdowns, but the term speedup is also kept for consistency. There are also the RV32I and \emph{max(IM, IF)} series. The latter represents the maximum performance between the fixed specifications RV32IM and RV32IF per individual run.

When selecting the (50,0)-cycle latency configuration, it can approach selecting the best extension per benchmark (\emph{max(IM, IF)} series) with an average performance at around 71\% of RV32IMF. It also exceeds the \emph{max(IM, IF)} performance in benchmarks like \emph{st} and \emph{wikisort}, where the use of ``F'' instructions is used more sporadically. Over a fixed baseline, when considering both the benchmarks classes ``improved by both F and M'' and ``improved by M'' (latter not in figure \ref{fireco}), a (50,0)-cycle latency configuration is 2.46x, 1.4x and 3.62x faster than RV32IF, RV32IM and RV32I respectively.

When comparing the versions for without and with a hit latency (lighter colours in figure \ref{fireco}), there is a considerable performance degradation at the higher values. For instance, for 250-cycle misses and 16-cycle hits, the approach performs similarly to featuring no instructions from ``M'' and ``F'', at 20\% of the RV32IMF performance. However, targeting a 0-to-few-cycle observable hit latency in future implementations, such as with fast FPGAs or more pipelining, seems to provide promising performance. This includes the (50,4)-cycle combination, which updates the above comparison of (50,0) to a speedup of 1.75x, 1.05x and 2.7x  over RV32IF, RV32IM and RV32I respectively. %

A general conclusion with regards to the latencies is that there is a sensible point where the approach is still helpful. %
For each workload there could be detailed curves with smaller latency intervals than the indicative values here. A similar argument can be made for the number of slots and other attributes. Though, analyzing specific points %
would be less significant, as this would relate more directly to the specification of the FPGAs and the core. %

\subsection{Multi-program}\label{miben}%

The effects of multi-processing are studied with the help of an operating system.  %
FreeRTOS \cite{barry2009freertos},  a real-time operating system, was selected to provide a minimal framework allowing experimentation with a task scheduler. %
A single binary is obtained, containing both FreeRTOS task scheduler and the benchmarks as threads. This is run as a bare-metal application by the adopted softcore %
to study the effects of context switching under our proposal for multi-programming. The main modification to FreeRTOS was the porting of the context-switching routine to support the ``F'' extension in our platform.

A periodic interrupt is set by its task scheduler, responsible for context-switching. The FreeRTOS scheduler enforces a round-robin priority between the tasks (benchmarks). A pair of benchmarks are run through two independent infinite loops, and once one of them does a certain number of iterations, %
the operating system terminates. 

Following the benchmark classification of section \ref{fispe}, the category that is not improved by ``F'' or ``M'' (``insensitive'') is not considered. The studied pairs are combinations between two of the five benchmarks that are improved by ``F'' and ``M'' (totalling 10) %
and combinations between one from the latter category with one from the eight benchmarks that are only improved by ``M'' (totalling 40). %

\begin{figure}[h!]
\centering
\includegraphics[width=1\textwidth, trim=0 0 0 0]{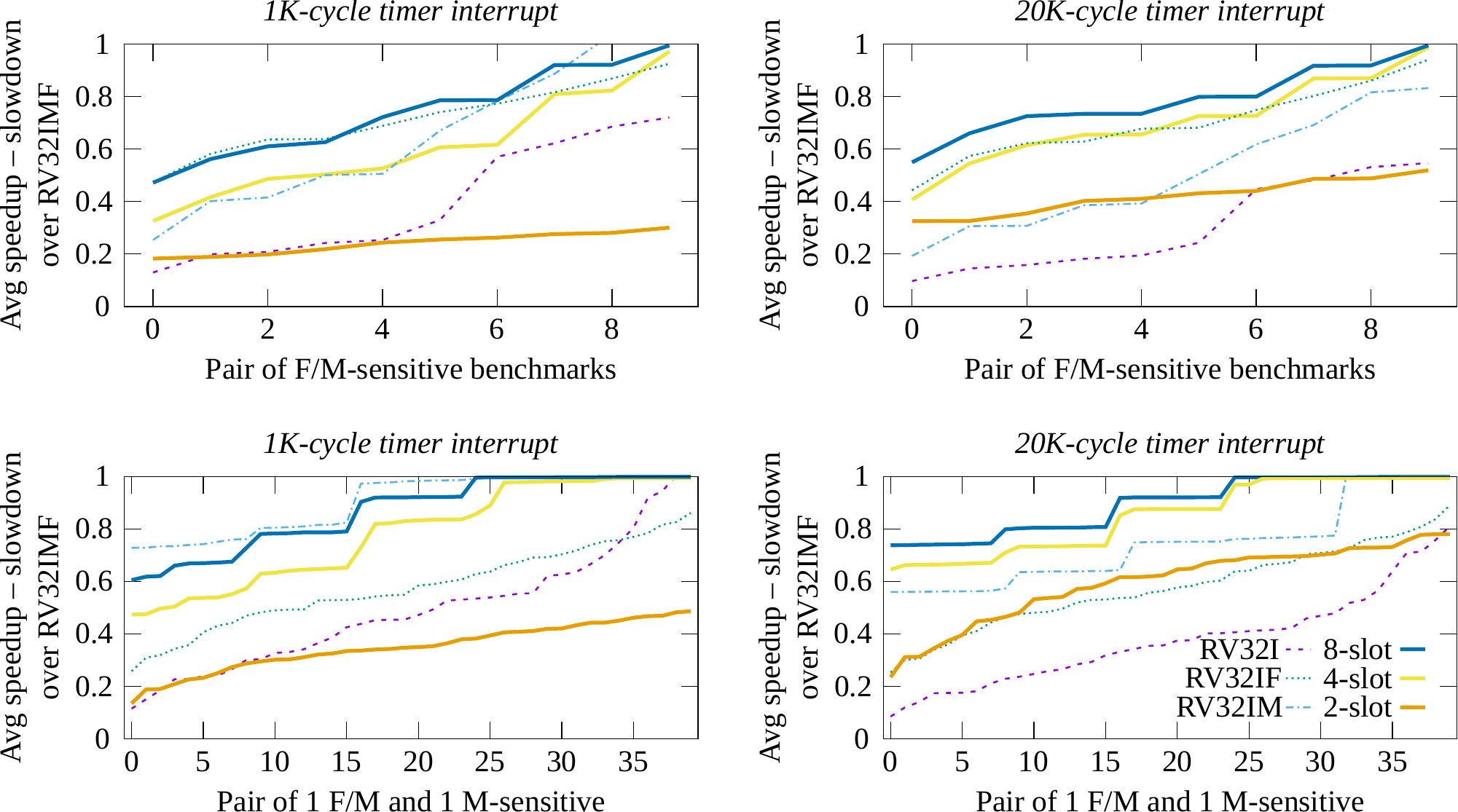}
\caption{Multi-programming using the reconfigurable approach with 2, 4 or 8 slots versus subsets of RV32IMF, under different %
scheduler timings. All series are sorted individually.}\label{multitf}
\end{figure}

Figure \ref{multitf} presents the results of this experiment with a 50-cycle miss latency (no hit latency) from the single-program experiments, as well as with variations of it for a different number of slots (2 and 8). The latter variations are added to elaborate on the slot interaction %
with this multi-program case, %
as the competitiveness between the slots is increased. The y-axes are the average speedups for each of the paired benchmarks over %
their corresponding runtimes with RV32IMF.%

The left plots in figure \ref{multitf} use binaries compiled for a 1000-cycle (1K) timer interrupt for context-switching, while the right plots present the results for a 20-fold increase in the timer interrupt delay. With the shorter 1K-cycle delay, all runtimes increase due to the additional instructions coming from the interrupt handler of the operating system. However, due to the different instruction distributions amongst the benchmarks, this also increases the instruction slot misses, hence the 20K-cycle versions improve the speedups of the reconfigurable approach. %
For instance, the average speedup of 4-slot series improves from 0.62 to 0.71 (i.e. from 38\% to 29\% slowdown) for the top selection of pairs, and from 0.82 to 0.9 for the benchmark pairs on the bottom of the figure. %

One observation when combining the benchmarks of the same class (figure \ref{multitf} top) %
is that the reconfigurable approach remains at the similar levels of performance degradation as with the last section (single-program). For instance, the average speedup for the 4-slot with 50-cycle reconfiguration and a 20K-cycle timer is 0.71, while the last section's corresponding average was also around 0.71. %

From figure \ref{multitf} (bottom right) we can see that the potential of reconfiguration is relatively higher when combining benchmarks with different extension preferences. The average speedup over RV32IMF for the 2-slot, 4-slot and 8-slot approaches is 0.62, 0.9 and 0.94 respectively, under 20K-cycle interrupts.

The proposed reconfigurable approach is shown to be more well-rounded than fixed extensions. %
For example, RV32IF performs significantly better than RV32IM in the pairs of the upper half of figure \ref{multitf}, but this is reversed for the pairs of the lower part. When considering all 50 of the aforementioned benchmark combinations for 20K cycle interrupts, the 4-slot version is 3.39x, 1.48x and 2.04x faster on average when compared to RV32I, RV32IM and RV32IF respectively, at an average of 0.82x the performance of RV32IMF. %
Finally, fine-tuning the operating system's scheduler parameters 
could be a cheap but necessary step to fully take advantage of the proposed computer architecture.

\section{Feasibility}\label{feas}

It is also important to comment on the readiness of current %
technologies to support such fast reconfiguration in future SoCs.
The main evaluation refrained from elaborating on this aspect to enable a discussion %
through system effects.

\subsection{Reconfiguration Latency Representativeness}\label{reconlat}

In order to demonstrate that future CPUs which feature FPGAs as functional units can be reprogrammed under a latency of the order of magnitude studied in  sections \ref{siben} and \ref{miben}, we present an example fast-reconfigurable FPGA architecture and prototype it in simulation.

The modelled FPGA is based on a traditional FPGA fabric layout but directly exposes a wide configuration bus which can be loaded from a wide bitstream cache. In contrast, typical FPGA architectures such as %
UltraScale+ constrain the reconfiguration port width to 32 bits \cite{xilinxrec}. 
The test designs for the FPGA were based on the RISC-V bit manipulation extension \cite{rvb}, %
including \emph{clmul} (carry-less multiply) and \emph{bextdep} (bit extract and deposition). %

The FPGA is modelled inside nextpnr \cite{shah} using the viaduct plugin framework for architectures, with a Verilog simulation model to confirm that bitstreams can be loaded in the target latency and function correctly. 
There are two connections between the FPGA fabric and the CPU. %
A wide configuration bus based on the Pico Co-Processor Interface (PCPI) from PicoRV32 \cite{wolf2019picorv32} is loaded through an L1 cache. Once the FPGA is configured, the fabric itself can also receive instruction operands and source register values; and returns a destination value after some cycles. %
This approach also enables partial instruction decoding; so one bitstream could implement multiple related instructions.

A series of optimisations are applied to the architecture to minimise the configuration array size (and hence cache size and configuration port width) and reconfiguration latency. %
The first relates to the removal of features less likely to be useful for this application, such as block RAM (BRAM) for storing large states. %
The inclusion of DSPs is not explored, %
though this could further reduce the configuration state by avoiding the use of fabric resources e.g. for multipliers. 

\begin{figure}[h!]
\centering
\begin{minipage}{.5\textwidth}
\centering
\begin{tikzpicture}[every node/.style={scale=0.92}]
 \begin{axis} [height=3.3cm, width=0.9\textwidth,
    xbar = 0.04cm,
    bar width = 6pt,
    xmin = 0, xmax = 200,
    ymin = -0.5, ymax = 2.5,
    ytick = {0,1,2}, %
    yticklabels = {4-LUT, 5-LUT, 6-LUT},
    xlabel= Bitstream size (K bits), %
    axis x line*=bottom, %
]
\addplot coordinates {(91,0) (114,1) (149,2)};%
\end{axis}
\begin{axis}[height=3.3cm, width=0.9\textwidth,
  axis x line*=top, axis y line=none,%
  xmin=0, xmax=4017,
  xtick =       {0,800,1600,2400,3200,4000},
  xticklabels = {0,800,1600,2400,3200,4000},
  xlabel=Configuration width,
  xticklabel pos=top,xlabel near ticks, ]
\addplot [opacity=0] coordinates {(0,0)};
\end{axis}
\end{tikzpicture}
\caption{LUT type versus bitstream size %
}\label{lutfig}
\vspace{-0.3em}
\end{minipage}%
\begin{minipage}{.5\textwidth}
\centering
\includegraphics[width=0.95\textwidth]{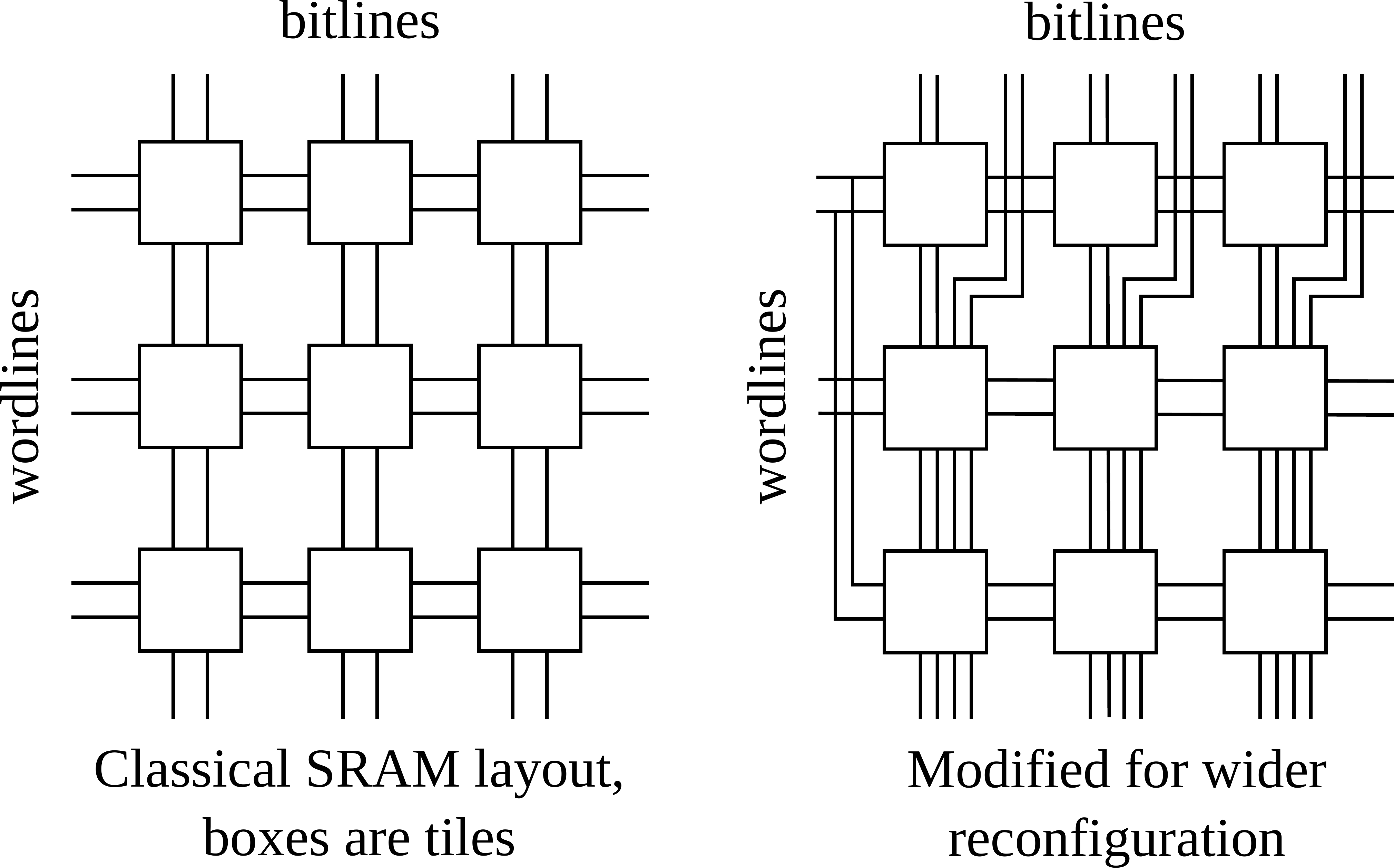}
\caption{Modification of SRAM %
FPGAs %
}\label{bitworldlinet}

\end{minipage}%
\end{figure}

An optimisation relates to the type of the look-up tables (LUTs), which are basic building blocks in FPGAs. 4-LUTs (i.e. with 4 inputs totalling 16 entries) are used rather than 6-LUTs. In this way the size of the configuration information %
is reduced; full instead of one-hot muxes is used for the routing; and the number of routing resources is generally minimised whilst keeping target designs routable. LUT permutation and route-throughs in place and route were used to partially compensate for the latter. The benefit of 4-LUTs in this context is shown with the  %
experiment of figure \ref{lutfig}, that determines the minimum configuration FPGA array size necessary to implement the \emph{bextdep} benchmark. %
Future work includes further optimising internal architecture, such as with fracturable LUTs.

When configured for 1680 LUTs, the bitstreams are a total of 91 kbits, requiring a 1824-bit wide configuration port for a 50-cycle reconfiguration latency. This is within the reasonable range of wide datapaths %
(see section \ref{bsds}), and it could be reduced at the expense of latency. Similarly, a 250-cycle latency (that still benefited some applications) would only require a 365-bit-wide port.

A necessary architectural change %
was to keep the entire configuration data path equal to the number of bitlines, rather than narrowing to an 8-bit or 32-bit external port or memory mapped configuration interface. This can then be loaded at full rate, in the target number of cycles, directly from the bitstream cache.

FPGAs generally use static RAM (SRAM) cells to store the configuration bits; and a word/bit line architecture to configure them. Architectures typically have a similar number of word and bitlines to ease routing. However, this would generally lead to unacceptably high configuration latencies for this application. 
 Reducing the configuration latency requires more bitlines and fewer wordlines -- the number of wordlines being equal to the latency, all things being equal. A diagrammatic example of the implication of increasing wordlines to reduce latency is shown in figure \ref{bitworldlinet}, simplified to %
few tiles and word/bitlines (showing only four configuration bits per tile, rather than a typical value of about a thousand).

This prototype uses a chain of shift registers to store the configuration bits. A configuration word is being shifted through the chain each configuration cycle (the chain is 50 deep, 1824 wide). This is for brevity, but has not challenged the routability of the test case. %
The operating frequency aspect is left as future work, but does not seem prohibitive at the moment, given the margins for a hit latency (section \ref{siben}) and reports for instruction-like tasks operating in the GHz range \cite{speedcore}. %

\subsection{Bitstream Cache Dimensions}\label{bsds}

To better understand the bitstream cache requirements for high-end processors, a separate study is conducted on a commercial x86 platform with a higher instruction variety (such as with vector 
instructions). %
By using dynamic binary instrumentation (DBI), %
this section explores %
the spatial needs and temporal localities %
with respect to the bitstream usage %
by comparing it to the traditional instruction and memory usage. 

The study of cache size requirements would normally involve measuring the \emph{working set} by simulating caches of different sizes and levels and pinpointing the size where the miss rate declines sharply. %
However, this could use assumptions relating to data and instructions, such as about the longevity of the working set (benefiting from multiple cache levels) and the access pattern (the notion of working set implies certain access distributions in space and time).

Through a custom Intel Pin \cite{luk2005pin} tool, on every dynamic instruction call, a routine updates a series of data structures for statistics on the opcodes, instruction pointers (IPs) and memory locations (where applicable). Each opcode is perceived as a separate bitstream. This represents the worst case to avoid specialisation in the observations, since it also includes control flow and data movement instructions, which are expected to be the most frequent \cite{akshintala2019x86,chang2008exploiting}, as well as for similar instructions that could be grouped together. %
A mask is applied to ignore the last 6 bits for a 64-byte-granularity in data and instruction blocks, which is commonly found in today's x86 systems. %

The data structures inside the Pin tool are mainly hashsets that provide the number of unique opcodes, instruction and data blocks. On every \(n\in \mathbb{N} \) number of instructions, %
the 3 corresponding sets are cleared and their cardinality is saved in lists (implemented as maps of \(<\)cardinality, occurrence\(>\) pairs to conserve memory). This provides the distribution of compulsory miss cardinalities occurring in the specified periods of time (measured in instructions) for each of the opcode, instruction and memory cache blocks. For the instructions and data, the algorithm's input would represent the stream observed right before the L1 instruction and L1 data caches. Though, the spatiotemporal locality scope of this experiment extends beyond the L1 caches. For the opcodes, %
this stream is considered to be observed from  the bitstream disambiguator. %

The benchmark suite selection for the single-program experiment is the single-core part of Geekbench 5. This is a series of 21 compute-intensive benchmarks ranging from encryption to machine learning one by one. Here, the same instance of the Intel Pin tool is used for the entirety of all Geekbench benchmarks.

\begin{figure}[h!]

\centering
\begin{minipage}{.5\textwidth}

\centering
\includegraphics[width=1\textwidth, trim=0 5 0 25]{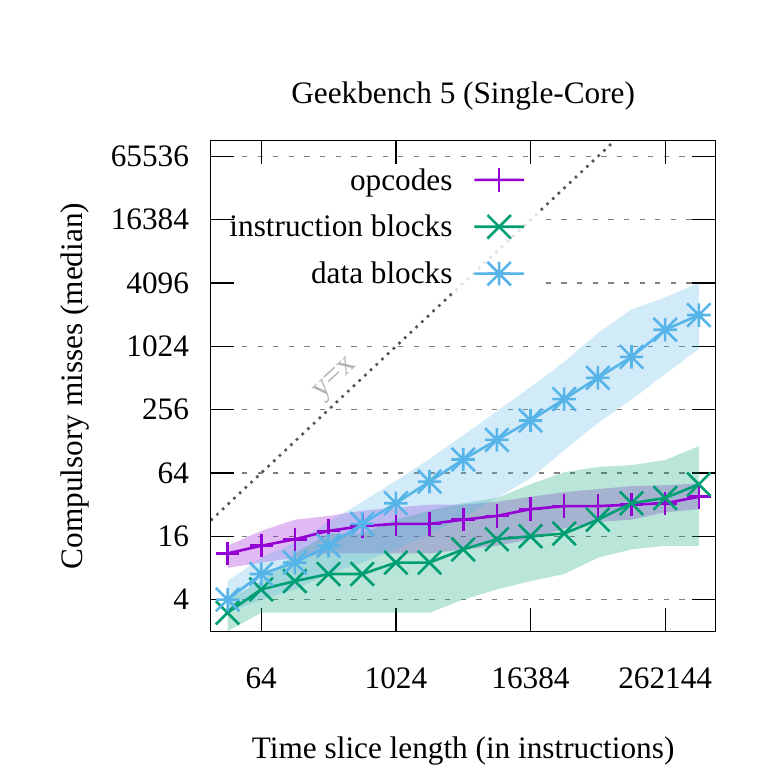}
\caption{Reuse behaviour in single-program%
}\label{opcodef}

\end{minipage}%
\begin{minipage}{.5\textwidth}
\centering
\includegraphics[width=1\textwidth, trim=0 5 0 25]{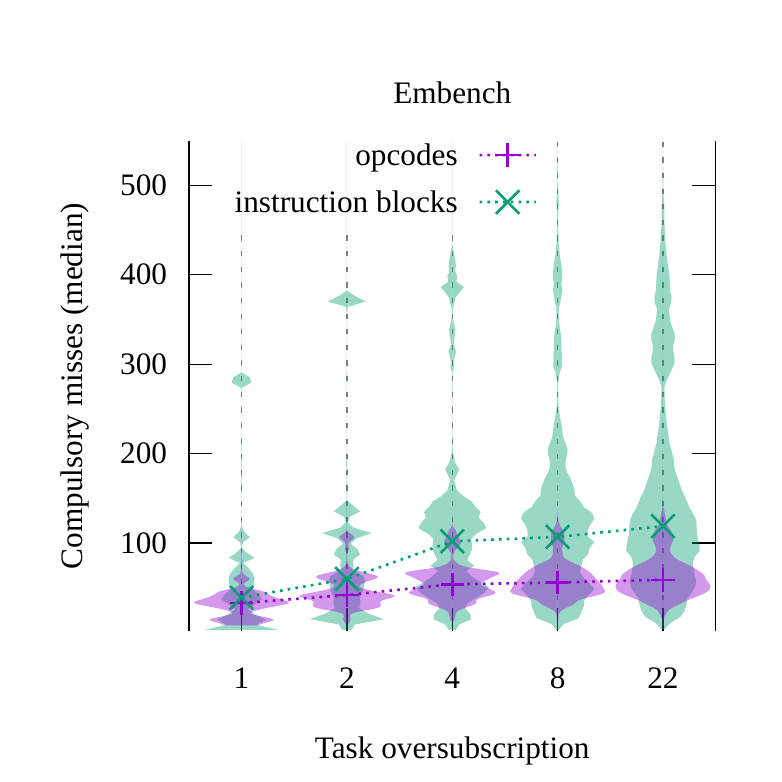}
\caption{Reuse behaviour in multi-program}\label{violinf}
\end{minipage}
\end{figure}

These results are illustrated in figure \ref{opcodef}. The \(x\)-axis summarises the time period the hashsets are collecting information for, and are used to observe temporal locality. The \(y\)-axis shows the observed median cardinalities for each hashset, and represents the compulsory misses for each type of cache block (bitstream for opcode, instruction for IPs and data for data addresses). The shaded regions underneath %
show the %
lower and upper quartiles of the cardinalities in each corresponding list of hashsets. As shown, the opcode count starts from below 16 for the shorter time slices, while peaking at below 64 for the longer time slices.

Another observation is that the opcode series is "flatter" than the other curves, meaning that there is higher reuse of opcodes than instruction and memory blocks. It is also longer lasting. This could have been conjectured, but the relationship between the 3 types of reuse is a result of multiple factors.
For instance, each instruction and data block already covers multiple locations (64-byte granularity), whose reuse also depends on the memory access pattern. From the opcode's perspective, CISC ISAs like x86-64 include a rather high number, as with Intel Pin's catalogue of over 8000 entries. %
The fact that the instruction and data series have a steeper upwards slope can also justify the need for multi-level cache hierarchies, as their requirements grow more rapidly with time. %

For the multi-program study, the same Pin tool (but with added thread safety mechanisms) is used on the adapted Embench suite for %
section \ref{miben}. This time each benchmark is run as a thread using \emph{pthreads} in Linux/x86\_64 and is pinned down to the same core for oversubscription. The benchmark suite is compiled as a single binary, and with the \emph{-march=native} flag to promote vectorisation. 
The idea of oversubscription here is to attempt increasing opcode demand, similar to the operating system's frequent migration of hundreds of tasks. %

The results of the multi-program experiment is shown in figure \ref{violinf}, where the median compulsory misses are measured for different amounts of task oversubscription to a single core. The time window size is fixed to \(32768\) instructions. Superimposed to the %
medians are violin plots, which are used to visually provide more detailed distribution information than percentiles. The data behaviour was fairly similar to the instruction blocks in this instance, hence its omission for readability. %
As expected, there is higher reuse of opcodes than instructions among the different tasks when oversubscribed. For example, when all 22 tasks are run simultaneously, the median and maximum opcode miss count reaches 59 and 152 respectively, while the respective numbers for instructions are 119 and 672. 

These numbers show that the bitstream cache is feasible with today's technology on SoCs. Specifically, a 64-block bitstream cache is shown to be enough for relatively long periods of time in both the single-program and multi-program experiments. This totals 768 KB of SRAM when using 12 KB bitstreams, being inline with the example FPGA architecture of section \ref{reconlat}. %
By observing recent processor trends that feature, for example, up to 256 MB L3 caches  \cite{suggs2020amd}, the sub-MB size requirement is in the L2 territory. Additionally, a sub 2048-bit datapath that is demonstrated in section \ref{reconlat} is only needed between this cache and the instruction disambiguator, as the expected latency profile of the bitstream cache makes progressively-loaded bitstream blocks meaningful, even with 128/256-bit datapaths to L2.
Given the oversubscription experiment results and the read-only nature of the bitstreams from the FPGA's view, future multicores would also benefit from sharing of the bitstream cache(s). 

These conclusions are drawn with the worst case approach in mind, such as by associating compulsory misses %
with the desirable cache size, and by not classifying opcodes into groups. A fraction of the reported desirable bitstream cache could still benefit future high-end CPUs with FPGAs working as instructions.

\section{Related Work}\label{rlw} 
Earlier research focused on using FPGAs as a functional unit. Garp \cite{hauser1997garp} targets embedded processors without multi-processing support, but it introduces the idea of combining a bitstream alongside the process binary. It does not make FPGAs transparent, as it requires configuration instructions. DISC \cite{wirthlin1995disc} is an earlier work that elaborates on reconfiguration in a similar context. Its instruction decoder is similar to the proposed instruction disambiguator by using caching. It is not a general-purpose computer architecture, as the processor has a separate ISA from the host processor. %
Chimaera \cite{ye2000chimaera} provides a reconfigurable array to dynamically load FPGA-based instruction implementations. This is somewhat reminiscent of the proposed bitstream cache, but 
only supports specially-compiled software. Architectures like CCA \cite{clark2005architecture} and RISPP \cite{bauer2009rispp} aimed to improve the adaptability of embedded systems by providing a set of specialised functional units that can be dynamically selected at run-time. The latter does not involve FPGAs. %

FABulous \cite{koch2021fabulous} is an open-source framework for integrating FPGAs into an ASIC. One of its applications is for the implementation of eFPGAs, also for the purposes of extending hardened cores. A RISC-V SoC with eFPGAs is presented as a use case. %
Related research studied the integration of SIMD units \cite{ordaz2018soft}, but the insights were platform-related, such as with regards to Xilinx' partial reconfiguration.
The custom instruction usage is limited to specialised kernels, and concepts like context-switching are not studied.

\section{Conclusions}\label{concs}
The FPGA-extended modified Harvard architecture %
can be used to transparently fetch standardised ISA extensions or custom instructions through the computer's memory hierarchy. The disambiguator unit works as an L0 cache for the FPGA slots and requests and multiplexes the bitstreams and instructions to reconfigurable regions. The evaluation showed promising results%
, generally surpassing the performance of a core with a constrained extension subset. %
The operating system in such computers is shown to benefit from longer times between context-switches to compensate for the reconfiguration time. Finally, %
a low reconfiguration latency is deemed necessary for the efficiency of the proposal, and our feasibility study finds this possible by mainly using existing FPGA building blocks and a cache with appropriate dimensions for providing the bitstreams. %

\subsubsection{Acknowledgement}
\small 
We would like to thank Anuj Vaishnav for his feedback on an earlier version. The second author mainly contributed with the section \ref{reconlat}.

\bibliographystyle{IEEEtranS}
{
\bibliography{refs}}

\end{document}